# Enhancing Traffic Flow Efficiency through an Innovative Decentralized Traffic Control Based on Traffic Bottlenecks

Nimrod Serok[1], Shlomo Havlin[2], and Efrat Blumenfeld Lieberthal[1*]

[1]Azrieli School of Architecture, Tel Aviv University, Tel Aviv 6997801, Israel
[2]Department of Physics, Bar-Ilan University, Ramat Gan 52900, Israel
*Corresponding author: efratbl@tauex.tau.ac.il

**Background**

Increased traffic demand has caused major congestion challenges (Ramezani et al., 2017) which result in billions of Euros that are spent yearly on external transport costs for the European Union (EU), congestion costs, and import of fossil fuels. Given this context, the EU has embarked on an initiative to integrate transportation, health, and environment concepts into sustainable and healthy mobility (Mihăiţă et al., 2018). While new infrastructure might have solved some of the problems, it is not done due to financial costs, environmental considerations, the diminishing availability of urban space, and due to fluctuations in demand during the day (Ramezani et al., 2017).

The advancement of intelligent transportation system (ITS) technologies, incorporating novel monitoring paradigms and computational systems, supports the real-time acquisition of traffic states and the execution of traffic-responsive signal control schemes. ITS provides an alternative to static, less affective approaches, under oversaturated conditions. However, while these cutting-edge technologies may reduce traffic congestion in localized, small-scale areas, achieving optimal traffic-responsive signal control across large scale networks remains a computationally challenging task. From a mathematical standpoint, this is classified as an NP-hard problem, meaning that determining the optimal solution is not possible on a real-world scale within real-time constraints (Ramezani et al., 2017). The optimization of traffic light control presents a convoluted problem, as it seeks to optimize a "self-organizing" system. A self-organizing system describes a complex arrangement where the ongoing interaction between its elements leads to the emergence of a cohesive global order, not governed by a single or a limited number of forces or elements. It is a result of the interactions between numerous elements, creating feedback mechanisms that effectively govern the system. Therefore, adopting a dynamic optimization process as an approach to solving self-organizing systems is useful and allows a dynamic adaptation to unanticipated changes through local element interactions (Cools et al., 2013).

Extensive research has been conducted on the design and optimization of isolated or coordinated signals that proactively address congestion within urban networks. Generally, signal control has leveraged two types of control mechanisms: static and vehicle-actuated controls (Hamilton et al., 2013). Static control, also referred to as a "fixed-time plan" focuses on optimizing the cycle time, the offset between adjacent intersections, and the allocation of green times to various directions within a cycle. This can be executed to establish a "green wave" wherein vehicles consistently arrive at intersections during the green cycle time, as demonstrated in (Gartner et al., 1975; Le et al., 2015). However, even optimized static control is unable to respond to dynamic situations arising in self-organized systems, thereby overlooking the intricacies of urban traffic.
Vehicle-actuated controls, on the other hand, use real-time, on-road detector data to enhance the temporal sequencing of signals on a per-cycle basis. Advanced implementations of this approach dynamically adjust traffic signal timing parameters (e.g., cycle length, phase split, and controller phase durations) in response to traffic conditions and changes. Vehicle-actuated controls can be categorized into two primary strategies:

1. Centralized methods - use a central computational unit that applies various approaches to optimize signal plans. These methods often frame the problem in a centralized manner, inherently limiting its scalability (Le et al., 2015). To date, the most popular traffic-responsive control method, employed in over 250 towns and cities, is SCOOT, which relies on a central computational unit to govern system-wide signal states (Thunig et al., 2019). Another method is "Hybrid systems" which combines fixed-time plans and vehicle-actuated controls; one important example is SCATS, which offers several predefined control plans for diverse traffic conditions but lacks the capability to optimize traffic parameters online (Le et al., 2015; Sims & Dobinson, 1980).

2. Decentralized solutions - try to reduce computational complexity and communication efforts between sensors and a central computational unit. These methods, also known as self-controlled signal methods, determine signal states locally without deep system knowledge. A major challenge for such solutions is ensuring the global stability of the system. To address this challenge, several methods have been developed (Lämmer & Helbing, 2010; Le et al., 2015), including system partitioning, back pressure calculation, self-organizing (SO) methods, genetic algorithms (GA), and reinforcement learning (RL). A critical element of local traffic optimization is the lack of negative interactions between neighboring intersections. Generally, optimization in one intersection may create suboptimal traffic flows at adjacent intersections due to spill-over effects. In fact, the main obstacle decentralized solutions need to overcome is mitigating negative interactions between neighboring intersections (Korecki & Helbing, 2022).

self-organized traffic light control is a promising concept for achieving both scalable (decentralized) solutions and mitigating negative interactions between neighboring intersections, as it fosters coordination among adjacent intersections (Cools et al., 2013; Korecki & Helbing, 2022). Decentralization offers tremendous benefits, including reduced computational costs, elimination of the need for centralized information aggregation, simplified implementation, and improves network robustness to failures (Le et al., 2015).

There are two main classes of coordination to achieve consequence traffic flow: A) active coordination, wherein each intersection's outcome is dependent on the outcome of others. An example of this can be found in studies that used multi-agent reinforcement learning (MARL) where the evaluation of the conducted policy for each agent depends not only on the observable surrounding but also on the policies of other agents (de Oliveira & Bazzan, 2009; Nowé et al., 2012; Wei, Xu, et al., 2019). And B) circumstantial coordination (also referred to as "loosely coupled"), wherein each intersection is independent in its decisions, but is affected **indirectly** by other intersections' decisions, as each intersection uses the data from sensors of nearby intersections which affect their decisions. Here, we focus on the latter which also contains several methodologies:

***Actuated traffic control*** – which is widely used in Germany. Due to their adaptive capabilities, actuated controllers often surpass fixed-time logic under various conditions (Branke et al., 2007). Vehicle-actuated approaches permit deviations in response based on the number of detected vehicles, enabling dynamic adaptation to current traffic patterns. When implemented at isolated intersections with sporadic traffic patterns, vehicle-actuated control can considerably reduce delays compared to fixed-time methods (Oertel & Wagner, 2011). This method configures minimum and maximum green light durations and within the range of these values, the phase is extended based on predefined restrictions. Given that actuated traffic control is a relatively straightforward mechanism with few configurable parameters and demonstrates robust performance in dynamic traffic conditions, it is frequently used as a benchmark when comparing traffic control methods (Aslani et al., 2017; Liang et al., 2018; Louw et al., 2022; Song et al., 2021; Tettamanti et al., 2017; Thunig et al., 2019).

***Self-organizing Traffic Lights (SOTL)*** - This approach calculates the number of vehicles approaching each intersection and determines the phase order and duration accordingly. The underlying idea includes accumulating "platoons" of vehicles that can travel together without stopping at subsequent intersections, as they receive priority based on their number, while in parallel creating gaps between platoons to allow crossing. This strategy has been proven to be very successful, surpassing state-of-the-art green wave methods that endeavor to optimize global traffic flow through traffic light synchronization and forming vehicle platoons that seldom need to halt (Cools et al., 2013). This concept was later elaborated upon by (Lämmer & Helbing, 2008).

***Back Pressure*** - which determines the subsequent phase based on the disparity in traffic load between roads leading into individual intersections and those leading out (Le et al., 2015; Varaiya, 2013; Wongpiromsarn et al., 2012; Zhang et al., 2012). In this approach, each intersection calculation employs only the queues directly leading to and from the intersection, ignoring data sharing between intersections and resulting in only circumstantial coordination. Although this method is simple to compute as it considers only single intersections, it has been found to outperform optimized static control in some scenarios (Le et al., 2015).

***Reinforcement Learning*** - Owing to the complexity of traffic light optimization, some recent studies have suggested employing machine learning approaches. These proposals advocate for iterative, neural-network-based learning (Korecki & Helbing, 2022; Wei, Chen, et al., 2019; Wei et al., 2018). Typically, each intersection calculates its phases based on incoming and outgoing lane data, which facilitates the scoring of previous decisions. The evaluation of each decision enables the machine to learn the optimal choices for various local situations. The underlying premise is that high scores will be awarded to decisions that effectively coordinate with neighboring intersections, without necessitating direct communication between them. Some of these solutions allow data transmission between intersections but not the sharing of each intersection's chosen phase (Korecki & Helbing, 2022).

**Partitioning -** here links are classified as congested or not based on their queue state. A cluster of consecutively congested links is created, and only its entry and exit points are managed to reduce congestion. In this case, data is communicated between intersections, but the decision made by each entry or exit intersection is not dependent on the choices of its neighboring intersections (Ji & Geroliminis, 2012; Ramezani et al., 2017; Sirmatel & Geroliminis, 2021).

**Genetic Algorithms (GA) or Evolutionary Algorithms (EA) -** for local optimization of intersections. This method utilizes a population-based approach, allowing the simultaneous exploration of multiple solutions in each iteration (Deb, 2011). EAs supports global search capabilities and can effectively balance solution quality and computational time for intricate organizational problems (Dong & Zhou, 2014). Various studies have demonstrated the suitability of EAs for multi-objective optimization problems, as the population-based meta-heuristic can generate a set of Pareto-optimal solutions in a single execution (Mihăiţă et al., 2018).

The "jungle-principle", introduced by Friedrich (2007) suggests that each intersection within a network controls traffic based on its local perspective. From a game-theoretical perspective, selfish optimization does not necessarily result in a system-wide optimum – it could even lead to poor performance, as demonstrated by the "tragedy of the commons" (Lämmer & Helbing, 2008). When traffic lights primarily respond to local demands, each intersection only reacts to traffic originating from neighboring intersections, making the dynamic interdependencies within the network unpredictable (Lämmer & Helbing, 2010). Recent studies show that urban bottlenecks generate a tree-like congestion street structure, with the cost of congestion being the sum of the cost of all the streets included in the tree (Serok et al., 2022). Using this method can help prioritize which congestion needs to be resolved first to improve the traffic on the entire network and not only locally. Here, we propose a decentralized, scalable, and circumstantial coordination method that employs "tree" shape identification and cost calculation for determining the priority of each phase (Zeng et al., n.d.).

As urban traffic is characterized by feedback loops, where congestion influences drivers' route choices, which in turn affect congestion levels and so on, evaluating a traffic control system requires simulation (or field experimentation). Given the high installation and operational costs associated with traffic-responsive signals, we present a simulation model which employs SUMO (Simulation of Urban Mobility) as the traffic simulator. SUMO is an open-source, microscopic, multimodal traffic simulator that enables users to assess the performance of specified traffic demand on a given road network. It is especially suitable for urban traffic simulation because it can execute optimized traffic distribution methods based on vehicle type or driver behavior. On top of that, it can update vehicles' routes in response to congestion or accidents, identifying optimal routing options to enhance urban sustainability and resilience. The Traffic Control Interface (TraCI) is another feature of SUMO that allows users to interact with the SUMO as a server. It includes a Python API which enables users to gather data from a traffic simulation or make modifications to the simulation (Koh et al., 2019; Liang et al., 2018; Song et al., 2021; Tettamanti et al., 2017). SUMO supports gap-based actuated traffic control (referred to as "actuated"), that extends traffic phases when a continuous stream of traffic is detected. The system transitions to the next phase after identifying a sufficient time gap between successive vehicles, resulting in improved green-time distribution among phases and adjusting cycle duration in response to dynamic traffic conditions (German Aerospace Center (DLR) and others, 2023).

In this paper, we present a new decentralized traffic light control approach aimed at optimizing traffic flows in real time, called the Tree Method. In the next section, we present the methodology on which we base this method and the simulation model we use to demonstrate its advantages. Next, we show the results of different simulations, and a comparison between the Tree Method and other benchmark methods. Lastly, we conclude by demonstrating the superiority of the Tree Method under specific conditions of high traffic loads.

**Methodology**

The proposed method, called the "Tree Method" is based on self-organized dynamic congestion trees. It is a "fixed cycle, fixed order" method, meaning the cycle length and the order of the phases are kept constant, but the duration assigned to each phase is determined at the beginning of each cycle. One of its significant advantages is the capability to selectively activate the Tree Method system automatically, limited to specific times or intersections. We find that targeted activation enhances control and efficiency in traffic management. Furthermore, this strategy bypasses the undesired consequences of erratic and unpredictable phase ordering, particularly within urban traffic environments. The introduction of unpredictability in phase sequencing may evoke driver frustration and potentially engender confusion, leading to unsafe and perilous driving behaviors. The main concept of our method is based on the principle of back-pressure caused by a temporal congested link (temporal traffic bottleneck), which results from the temporal gap between the supply and demand in each traffic light phase. This temporal bottleneck can cause a chain of congested streets on its upstream (a "tree"-shape congestion, see (Serok et al., 2022)N19N19 for elaboration), and affects distant streets and even streets of higher hierarchy. To identify and estimate the cost of each tree we follow (Serok et al., 2022) and create an adaptive method of a self-organized traffic control system. We use the Tree Method to calculate the cost of congested trees for each intersection at the end of each cycle and use this cost to define the duration in each phase for the next cycle. For that, we use a multi-stage methodology:

(1) *Network representation and pre-calculations* - In the context of a road network where a road approaches an intersection and diverges into multiple directions (such as right, straight, and left), the implementation of traffic signal phases is crucial to manage the flow of vehicles and prevent conflicts within the intersection area. To accurately align the duration of these phases with the actual demand, we use the road infrastructure to represent a network that allows the estimation of demand for each direction from every incoming road. Typically, drivers utilize all lanes on the same road based on their driving velocity, with faster drivers occupying the left lanes. As the vehicles approach the intersection, they switch to the appropriate lane corresponding to their intended direction. Thus, to effectively model the traffic dynamics, we divided the street segment between two intersections into several segments or links comprised of an initial segment, which extends from the upstream intersection to the point where the lanes separate into different directions. We also defined head links that correspond to a specific phase of the signal cycle, allowing traffic flow towards different directions. The urban area is therefore transformed into a network (Fig 1A), such that the nodes represent intersections or the merger of lanes connected to an intersection. The links represent the street segments between the nodes, where the head links typically correspond to specific phases of the traffic signal cycle and may be served by one or more existing signal phases. Alternatively, they can represent free turns, allowing vehicles to make turns in any direction during any signal phase. Body links refer to the actual street segments that reach the head links. As a result, the demand on a particular signal phase, considering the existing roads and traffic signal configuration, can be quantified as the **demand** on a set of links.

(2) *Tree calculation* - The cycle cost for each body link within the prescribed cycle time is derived from the time delay experienced on the street segment relative to the maximum flow rate. If the velocity on a body link exceeds the maximum flow velocity, the cost value is set to zero. Body links with head links leading to an intersection and having a non-zero cost (indicating congestion) are identified as Trunks, representing the roots of Congestion Trees. For each trunk, a Congestion Tree is recursively constructed, by adding the congested body links that lead to it, as well as the congested body links leading to those links, and so forth, until all leading body links are not congested. Thus, within a specific cycle, a body link can be a part of a few trees simultaneously, but it can only serve as a trunk of a single tree. To simplify the terminology, all body links currently participating in a tree, whether trunk or not, are referred to as branches. For each branch, we keep the number of trees in which the branch is involved. The cost contributed by each branch to its respective tree, known as branch cost, is calculated by dividing the segment's cost by the corresponding trees-counter. Once all trees have been evaluated, the cost of each tree is computed as the sum of all its branch costs. Consequently, at the conclusion of each cycle, the resulting trees are obtained along with their associated costs, measured in vehicle hours. This calculation is mathematically straightforward and can be performed in a decentralized manner, by facilitating data communication between neighboring street segments (Fig 1B).

(3) *Calculating the cost associated with each intersection phase*: at the end of each cycle, the model calculates for each intersection the cost for each phase in the last cycle. Each phase facilitates the movement of vehicles along specific body links and their corresponding head links. The cost of a body link $C_{ij}(t)$ is calculated for every cycle, relative to its cost on free-flow speed $U_f$. This cost represents the time it takes to cross a road (link) in comparison to the time it takes to cross this road in a maximal flow (calculated for each link), multiplied by the number of drivers who crossed the endpoint of this link at a specific time:

(1) $C_{ij}(t) = dist_{ij} * \left( \frac{1}{u_{ij}(t)} - \frac{1}{u_{qmax_{ij}}} \right) * \frac{q_{ij}(t) * l_{ij}}{\frac{60}{T}}$

Here, $dist_{ij}$ is the length of the link in km, $q_{ij}(t)$ is the current flow on the link, $u_{ij}(t)$ is the current speed on the link, $u_{qo_{ij}}$ is the speed when the flow is optimal, $T$ represents a measurement unit (in minutes) which corresponds to the traffic light cycle (1.5 minutes in this case) and $l_{ij}$ is the number of lanes in the link (i.e. the number of lanes in each street segment of the JT). According to this formula, if the body link is not congested its cost value is set to zero. This cost $C_{ij}(t)$ of the body link, is divided equally between all the trees that contain this body link. As explained, the cost of each tree is assigned to its trunk (Fig 1B).

For each head link, a weight between 0 and 1 is assigned, considering the number of vehicles that drove on it throughout the previous cycle. This weight is normalized with respect to other head links associated with the same body link but used during different phases of the current cycle. The cost of a phase is thus determined by summing the costs of its corresponding body links, with each cost being multiplied by the weight assigned to the phase's head links.

(4) *Calculation of Intersection Phase Durations*: at the end of each cycle, each intersection undergoes a process to determine the durations of each phase in the subsequent cycle. The cycle length and the order of phases remain fixed, resulting in a competitive allocation of durations among the phases. This process begins by calculating the cost of each phase as explained above. Next, we assess the required adjustments: as the weights assigned to the phases reflect their durations in the current cycle, the algorithm aims to balance the durations of the next cycle based on the performance of the current one. The necessary adjustment is determined by considering the overall cost. A small adjustment may suffice if the sum of the costs for the phases in the current cycle is relatively small, whereas a larger adjustment may be necessary if the cost is substantial. The available duration to be divided among the phases is determined, accounting for both the minimum duration per phase and the size of the adjustment required. Finally, available duration is divided among the phases according to their respective costs. This division ensures that the durations reflect the relative costs of the phases.

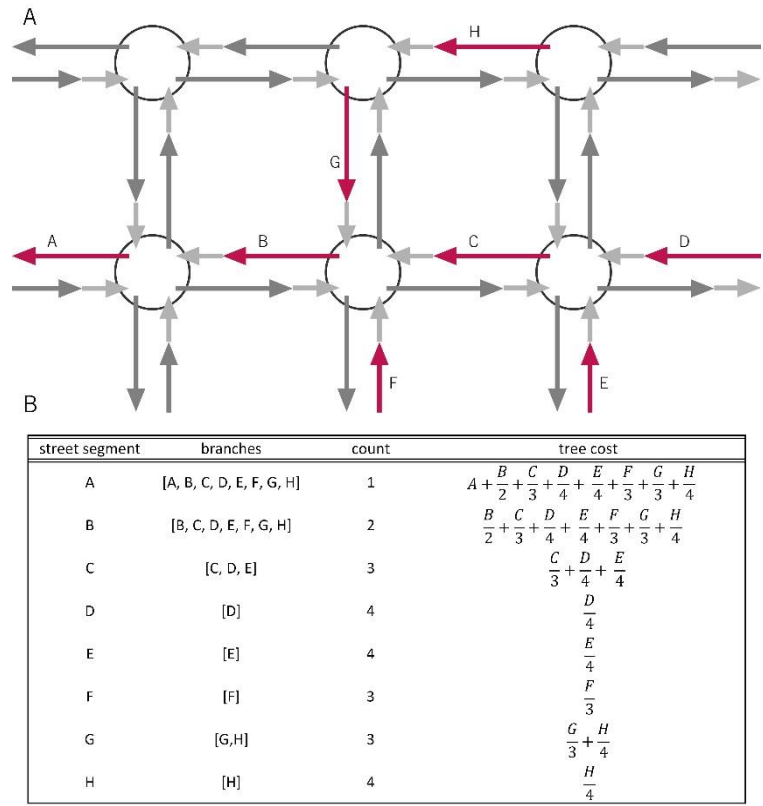

**Figure 1.** (A) The scheme demonstrates a current state of congestion in the traffic network, where red links represent congested body links, dark grey links represent non-congested body links, and light grey arrows represent head links. In this example, a total of 8 congested trees can be identified. (B) Each congested tree is depicted in this representation, with its trunk and corresponding cost which is determined by its branches. The branches represent the individual street segments that constitute the tree with the segment in the left column acting as its trunk. The count denotes the total number of unique trees in which this trunk is present, both as a trunk and as branches, and quantifies their contributions to the overall tree cost.

This computation process generates a situation where each occurrence at each node has a value that represents its impact on the system, whether it affects only a single node or creates congestion that affects subsequent nodes. When the supply-demand ratio creates a congestion tree, each relevant occurrence within this tree is assigned to its own subtree with its associated cost. Thus, each occurrence is treated as part of the same method. With congestion trees originating from different directions, each tree can receive priority at the merging point based on its cost, thereby minimizing the likelihood of a scenario where a minor congestion blocks a major congestion at a specific node. In our previous studies, we observed cases where a secondary street caused significant congestion on a main street. This approach significantly reduces the probability of such situations by allocating a central portion of the cost of the main street to the secondary street causing the congestion. This method also adapts well to changing demand conditions and accommodates fluctuations since it aligns the supply with the measured demand in each cycle.

Following the establishment of the model infrastructure, a series of simulations were conducted to assess the efficacy of the developed SODT method in comparison to alternative ones. These simulations were comprised of diverse scenarios aimed at evaluating the performance of the tested methods under varying conditions. The simulated experiments were conducted within a virtual city environment (Fig 2), with the resulting outcomes subsequently compared. The virtual environment is based on a real area for and traffic in Tel Aviv center, where we collected data on traffic volume and velocity as well as on the traffic light cycles. This data was utilized to construct an OD matrix that accurately reflects real-life traffic scenarios. This area was chosen as it represents an environment that experiences significant traffic congestions in opposite directions. This allows us to accurately examine the advantage of our method when facing complex urban traffic scenarios.

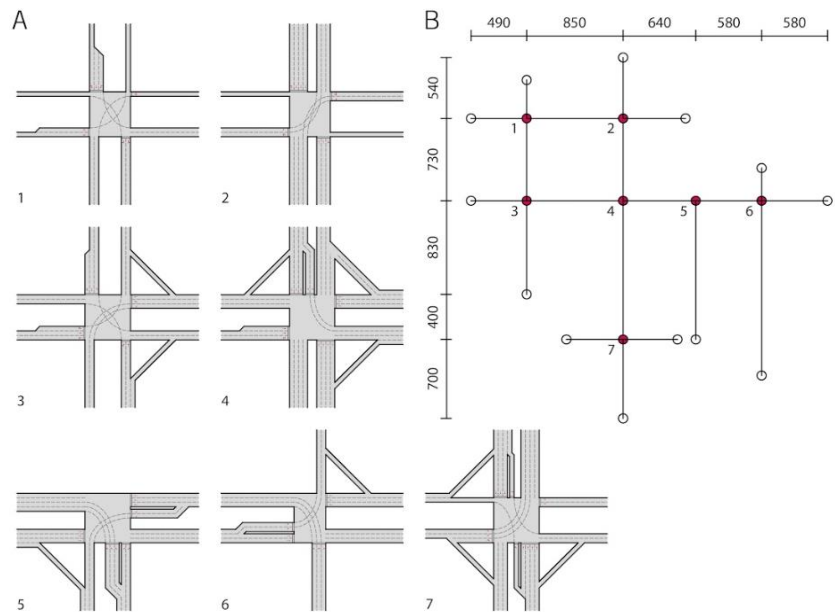

**Figure 2.** A) The virtual environment derived from a real area situated in the central region of Tel-Aviv, consisted of 7 junctions controlled by traffic lights. (B) The layout of these junctions, represented by red circles, along with the entrances and exits of the model, represented as open circles. The distances between these elements were carefully considered in the design of the experiments. Note that this environment served as the layout for all the experiments conducted in this study.

We ran four distinct experiments utilizing SUMO's TraCI engine for traffic-light control. The network structure and traffic light phases remained consistent for all four experiments, where each simulation represents a duration of two hours, and a predefined list of the number of vehicles per minute and their distribution in terms of origin and destination was employed for all simulations. Each scenario started by generating a unique vehicle list used for a particular experiment's configuration. Leveraging SUMO's TraCI engine, vehicles within the simulation had the ability to dynamically adjust their route choice based on prevailing traffic conditions, akin to real-time navigation applications such as Google Directions or Waze.

To compare our method with competing alternatives, we used identical traffic light phases with varying durations (according to each method), and fixed time cycle of 90 seconds across all the experiments and employed three other methods, each representing a different approach towards traffic-light control: (1) Fixed-time – where a fixed time was divided uniformly between all phases. (2) Random – where a random phase is chosen every 10 seconds. And (3) SUMO-actuated - A gap-based actuated method that is built-in SUMO. This method determines the next phase and its duration by analyzing virtual detectors on each lane. The specific parameters employed to configure this method are outlined in Table S1, and are considered a common practice benchmark within the field (Louw et al., 2022).

Each method was used for two OD matrices: the first one is a random matrix where the origin and destination for each vehicle are chosen randomly, and the other is based on real data of traffic volumes that were used to generate probabilities for the origins and destinations in the matrix which is also referred here as a realistic approach. Each vehicle that entered the simulation was assigned an origin and destination, based on these probabilities. Additionally, each method was used with two different traffic volumes – high traffic load (about 25,500 vehicles over 2 hours) and a moderate one (that corresponds to 75% of the high traffic volume).

**Results**

The main performance metrics employed for comparison in this study are the average travel time, measured in seconds, and the throughput, quantified as the number of vehicles over the entire simulation period. Both measurements were calculated using SUMO's TraCI engine based on only vehicles that successfully reached their destinations within the simulation time frame. We ran each experiment 20 times and found that the Tree Method indeed outperforms all alternatives method. Figure 3 and 4 (and Table S2) show the results of the average travel time and the throughput for the random OD matrix and data-based one correspondingly. For the runs that used the random OD matrix, with high traffic load, we found that in terms of the average travel time (based on all 20 runs), our method exhibits a reduction of 40% in comparison to the other methods. In terms of throughput, the results of our method also present a significant improvement of 35% compared to the other methods.

In the moderate load scenario, using the OD matrix, the tree method yielded an average increase in the throughput of over 19% and a reduction in the average travel time of over 43%. Notably, the absolute performance of the moderate load cases was marginally better than that of the high load cases across all traffic light control methods and measurements. To summarize, the Tree Method presented better results under random OD matrix for both high and moderate traffic load.

The SUMO-actuated method outperformed the Uniform and the Random methods, although it is worth mentioning that the Uniform method exhibited superior performance in terms of average driving time under high-load conditions. This observation suggests potential limitations of the actuated method when confronted with high loads originating from multiple directions simultaneously.

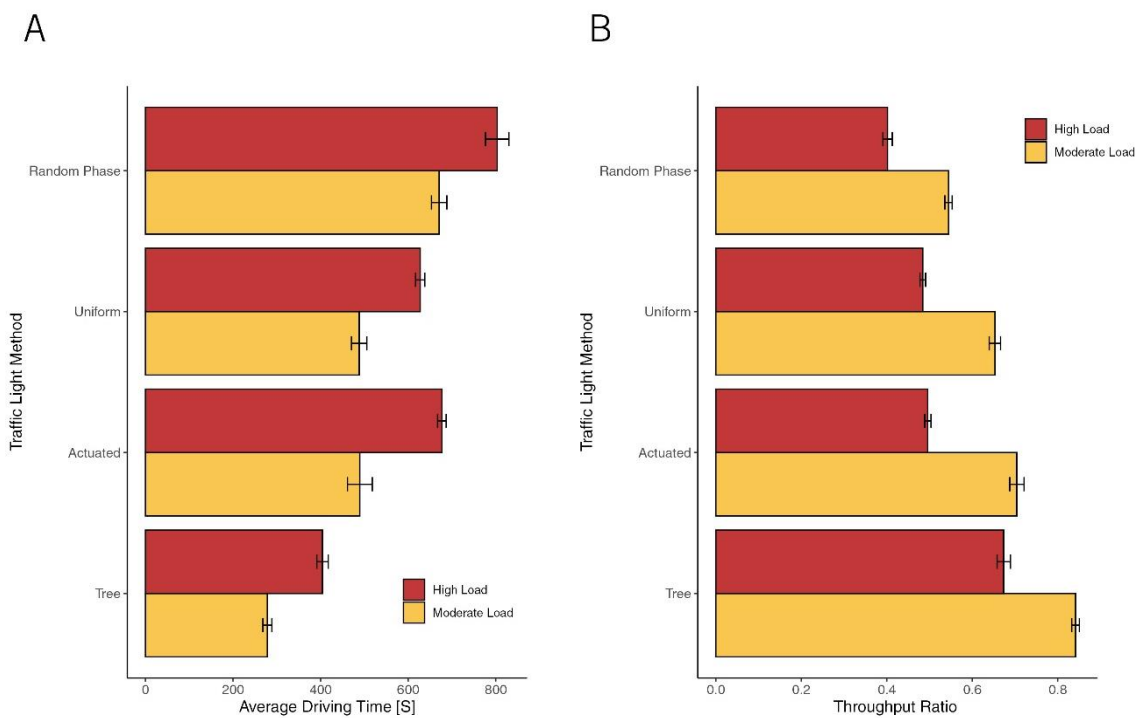

**Figure 3.** A comparison between the various traffic light control methods tested over 20 runs each on random OD scenarios with high traffic load (Red) and moderate traffic load (Yellow). (A) Average driving time (B) Throughput

While the utilization of random origin-destination simulations is a common practice for comparative analysis, it should be noted that such comparisons are theoretical and may deviate significantly from real-world scenarios.

A more realistic simulation approach considers incorporating measured load data from each entrance and exit point within the simulated area. The observed drawback of actuated methods when encountering heavy flows in opposing directions represents a complex challenge for decentralized approaches. Consequently, we present the results of the more realistic simulation where the presence of competing loaded flows is evident, aiming to capture the intricacies of such scenarios (Fig. 4).

The results of the simulations based on realistic OD matrix show that here too, the Tree Method achieves the best results for both measurements and both load levels. In comparison to the SUMO-actuated method, the throughput within the Tree Method presented better results by over 23% (on average) for the high load and over 7% for the moderate one. Additionally, the average travel time in the Tree Method was 43% less in comparison to the SUMO-actuated method for the high load and 21% less for the moderate one. Similarly, to the random OD matrix, the SUMO-actuated method was outperformed by the uniform method on the loaded cases. In these simulations, the Tree Method demonstrates similar results in comparison to the SUMO-actuated one on both measurements. In other words, when using real-life data with moderate traffic load, the Tree Method and the SUMO-actuated one presented relatively similar performance, whereas on the high traffic load simulations this improvement was doubled, suggesting a superiority of the Tree Method for cases of conflicting high traffic loads.

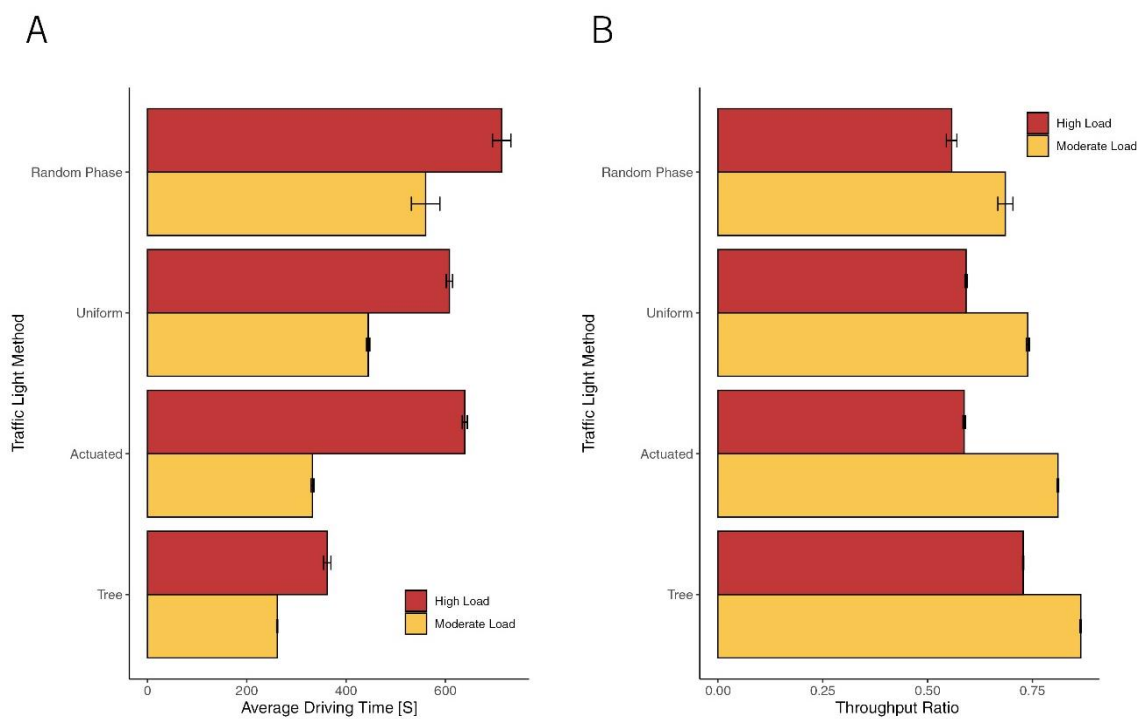

**Figure 4.** A comparison between the various traffic light control methods tested over 20 runs each on realistic OD scenarios with high traffic load (Red) and moderate traffic load (Yellow). (A) Average driving time (B) Throughput

The Tree Method Vs. the SUMO-actuated one

In figures 3 and 4, we examined four scenarios where we used moderate traffic, and high load traffic. On two different OD matrices: random and realistic one. Next we focus on the real-world high-traffic load scenario, which we consider highly relevant to real-life situations. As both the SUMO-actuated and Tree Method are dynamic scenarios we compare them here. Figure 5 provides a more detailed analysis of the high load scenario for these two methods, specifically zooming in on the driving time distribution of each run. It is evident that the cumulative distribution function (CDF) of each run in the Tree Method consistently performs better than the SUMO-actuated method for each driving time. The inset in Fig 5 presents the ratio between the cumulative distribution function (CDF) values of the Tree Method and the Sumo actuated method. It is evident that the Tree

Method outperforms the Sumo-actuated one for all driving times, but in particular for short driving times (above 5 minutes) where the ratio of the CDF values of these methods reaches 1.75. After 5 minutes (simulation time) 63% of the drivers in the Tree Method simulation have successfully reached their destination, while in the Sumo-actuated simulation only 36% of driver reached destination. This observation suggests that the Tree Method surpasses the SUMO-actuated method in a significant manner, not only in terms of average driving time but also for any driving time and each unique run.

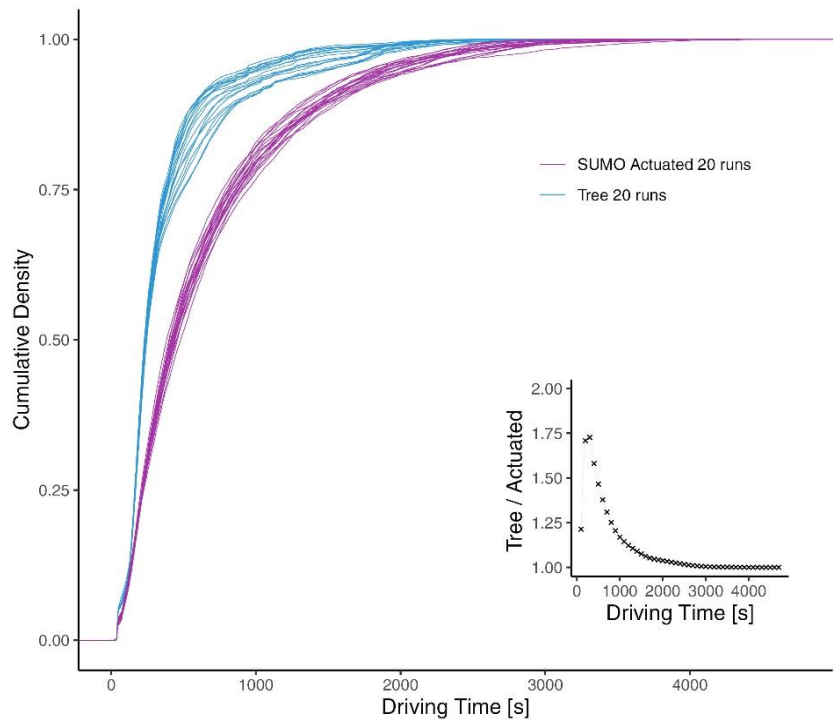

**Figure 5.** The cumulative density of the driving time for the realistic scenario. A comparison between 20 runs of the SUMO Actuated method and the Tree Method. In the inset, the cumulative density ratio between the two methods.

An additional detailed analysis focuses on the cumulative throughput in the simulation period. Each simulation run consisted of 80 traffic-light cycles, each lasting 90 seconds, spanning a total of two simulated hours. At the end of each cycle, the number of drivers who successfully reached their destinations was added to the cumulative throughput. Figure 6A presents the results of the throughput over time for 20 simulation runs using the tree method and 20 simulation runs using the SUMO-actuated method. Notably, all 40 simulation runs exhibit relatively linear and smooth curves without significant peaks representing high or low efficiency. This observation suggests that both traffic light methods demonstrate stability over time, with their results showing consistency regardless of the simulation duration. Additionally, the performance superiority of the Tree Method over the SUMO-actuated one is apparent in terms of throughput measurement. The Tree Method consistently achieves a higher number of drivers reaching their destinations in comparison to the SUMO-actuated method. Furthermore, the variations between different runs within the Tree Method are less pronounced compared to the Sumo-actuated ones. These findings highlight the enhanced effectiveness and reliability of the Tree Method in improving traffic flow and facilitating the successful completion of journeys by a greater number of drivers.

The Tree Method utilized in this study employs congestion-tree analysis, where each decision made during the traffic-light cycle has a continuous feedback loop that impacts the entire simulation. Analyzing the congestion trees that are created throughout the simulation provides insights into the internal workings of the Tree Method. Figure 6B illustrates the cumulative cost of the congested street in the simulation over time, which forms the basis of the tree analysis. As the goal of the tree method is to minimize this cost, it is not surprising to observe that the cumulative cost curves of all 20 runs in the Tree Method simulations are consistently and significantly lower than those of the SUMO-actuated ones. Furthermore, another noteworthy measurement is the average number of congested trees per simulation. Figure S1 depicts that the number of trees generated using the Tree Method is 59% of the number of congested trees created when applying the tree analysis to the SUMO-actuated method. This finding suggests that while the SUMO-actuated method deals with congestion that may not necessarily be the bottleneck of the jam, the Tree Method's inherent focus on these exact traffic jam bottlenecks enabling it to eliminate trees altogether instead of dividing them into smaller components.

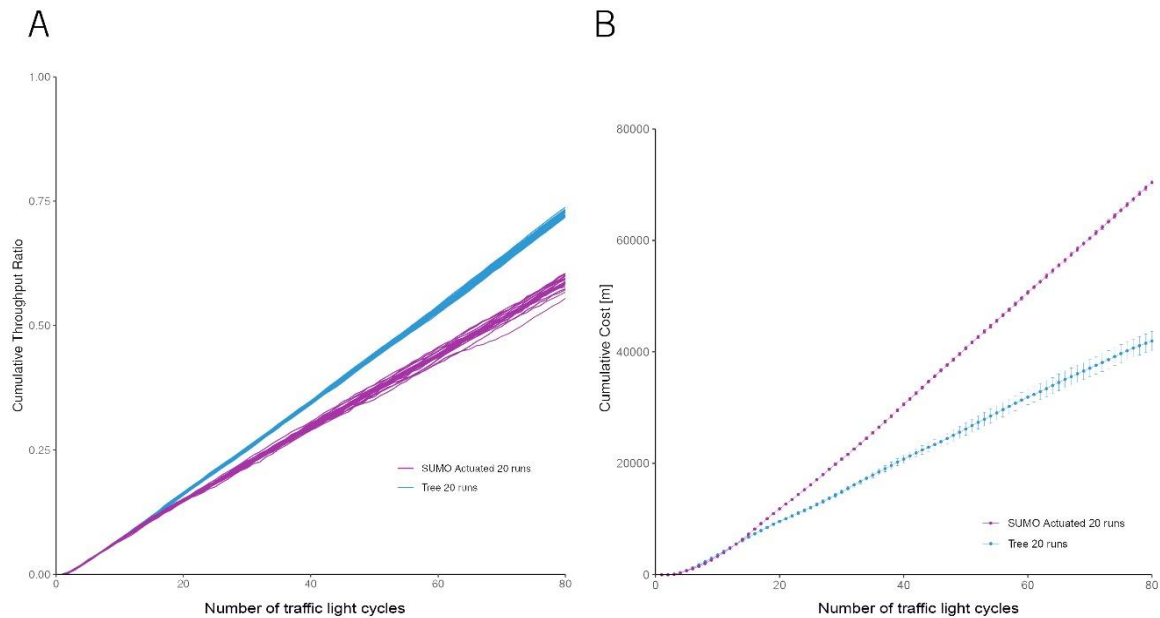

**Figure 6.** An analysis of the realistic scenario over time. A comparison between 20 runs of the SUMO- actuated method and the Tree Method. (A) Cumulative throughput over time (B) Average cumulative cost over time for 20 runs

**Discussion and Conclusions**

This paper introduces a decentralized traffic-responsive signal control method based on the identification of congestion-tree structures caused by traffic bottlenecks. By calculating the cost of each congested tree, this approach prioritizes bottlenecks based on their influence not only locally (i.e., on individual road segments) but also globally, meaning on the entire road network. The Tree Method determines the cost of each tree at the end of each cycle and allocates the duration of the next cycle's phases accordingly.

To compare adaptive traffic light control approaches, we conducted simulations using both abstract and realistic scenarios. The SUMO-actuated method served as the benchmark due to its simplicity of implementation and configuration, as well as its proven effectiveness in real-world traffic situations, particularly in Germany. However, the actuated method, which extends traffic phases when a continuous stream of traffic is detected, might not be the best solution when dealing with conflicting traffic flows that involve bottlenecks. This represents a significant challenge for decentralized traffic light control systems. Our simulation results support this understanding, with the SUMO-actuated method outperforming static methods in moderate-load simulations but being surpassed by the Uniform method in high-load simulations.

To address the challenge of conflicting traffic flows (i.e. traffic flows that complete on opposing cycle times in a specific phase of the traffic light), we proposed a new decentralized traffic light control methodology based on the identification of the congestion's bottleneck and its influence. The Tree Method consistently outperformed all other tested methods on all simulations. In the simulations of the realistic scenario (a realistic OD matrix based on real data) with high-load conditions, the Tree Method achieved an average increase in throughput of over 23% and a reduction in average travel time of over 43%. Further analyses of individual simulation runs and performance over time show that the Tree Method is superior compared to other methods not only in terms of average performance but also for the majority of drivers and throughout most of the time. By prioritizing traffic flows based on their global cost, rather than just their local one, the Tree Method accurately identifies the root cause of traffic-tree congestion and its impact along the traffic upstream. An additional advantage of this method is its straightforward analysis, facilitating real-time adjustments in each cycle and aligning with the inherent feedback dynamics of traffic flows. This characteristic makes this method highly suitable for implementation in real-world traffic control systems. Comparing moderate-load simulations to high-load simulations suggests that the advantages of the Tree Method become more prominent as the traffic load of competing flows increases. However, the nature of the Tree Method, which maintains the cycle duration while calculating traffic trees at the end of each cycle, enables its activation specifically when high-load competing traffic flows are detected and thus improving computing time over time.

In summary, the Tree method presented in this study is a decentralized traffic light control approach designed to optimize competing loaded traffic flows in real-time. When compared to the actuated traffic method, a commonly used benchmark (here based on the SUMO-actuated one), the Tree Method exhibited major improvements in both the throughput and average travel time measurements. The increasing integration of smart technologies, such as information and communication technology (ICT) and the Internet of Things (IoT), has resulted in advancements in queue length estimation, making it more feasible and accessible. Hence, considering its inherent simplicity, the Tree Method exhibits significant potential for application in real-world urban environments, offering the prospect of enhancing traffic flow within urban areas.

**Data availability**

For contractual reasons, we cannot make the empirical data from Google Direction available. However, all data from our analysis are available at GitHub: https://github.com/nimrodSerokTAU/bottlenecks-prioritization.

**Code availability**

The code of our analysis is also available at GitHub: https://github.com/nimrodSerokTAU/bottlenecksprioritization.

**Competing Interest Statement**

The authors declare no competing interests.

**Funding**

This work was supported by the European Union's Horizon 2020 research and innovation program (Grant Agreement 953783); City Center, TAU Research Center for Cities and Urbanism; and the Center for Innovative Transportation for supporting this research.

**Statement**

During the preparation of this work the authors used ChatGPT to perform language editing and spellcheck. After using this tool, the authors reviewed and edited the content as needed and take full responsibility for the content of the publication.

**Supplementary Information**

| Parameter | Value |
|---|---|
| Detector gap | 1 |
| Minimum duration (all phases) | 10 |
| Maximum duration (all phases) | 60 |
| Frequency | 300 |
| Max gap | 3 |
| Passing time | 10 |

**Table S1.** Configuration for gap-based actuated

|  | Average driving time [min] | | | | Throughput | | | |
|---|---|---|---|---|---|---|---|---|
|  | Rand moderate-load | Rand high-load | Realistic moderate-load | Realistic high-load | Rand moderate-load | Rand high-load | Realistic moderate-load | Realistic high-load |
| Random Phase | 671 | 803 | 561 | 714 | 0.54 | 0.40 | 0.69 | 0.56 |
| Uniform | 488 | 628 | 445 | 608 | 0.65 | 0.48 | 0.74 | 0.59 |
| Actuated | 490 | 677 | 332 | 639 | 0.70 | 0.50 | 0.81 | 0.59 |
| Tree | 278 | 404 | 262 | 362 | 0.84 | 0.67 | 0.87 | 0.73 |
| Tree / Actuated | 0.57 | 0.60 | 0.79 | 0.57 | 1.20 | 1.36 | 1.07 | 1.24 |

**Table S2.** Average driving time and throughput of different traffic signal control methods

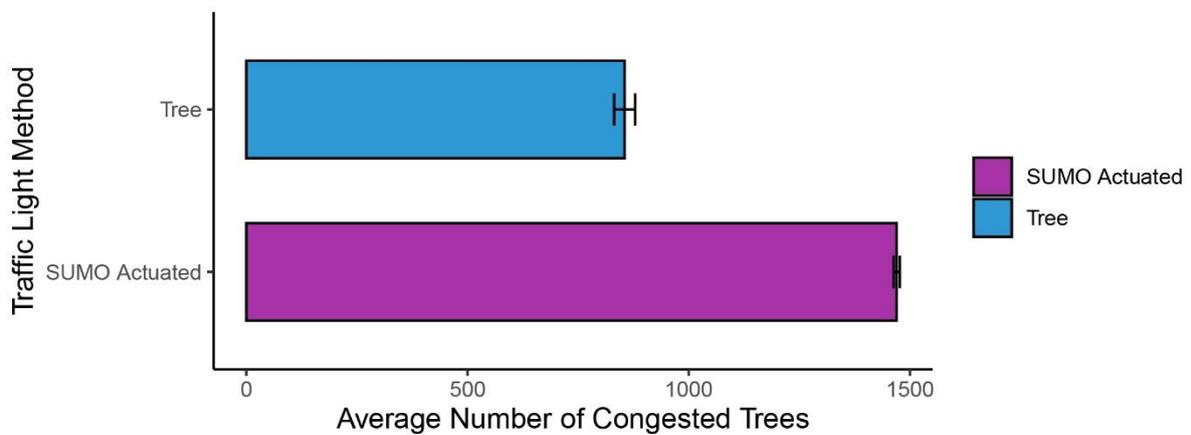

**Figure S1**. Average number of congested trees per simulation